**From Photons to Electrons: Accelerated Materials Discovery via Random Libraries and Automated Scanning Transmission Electron Microscopy**

Boris Slautin[1,*], Kamyar Barakati[1], Utkarsh Pratiush[1], Christopher D. Lowe[2], Catherine C. Bodinger[2], Brandi M. Cossairt[2], Mahshid Ahmadi[1], Austin Houston[1], Timur Bazhirov[3], Jaehyung Lee[4], Kamal Choudhary[4], Gerd Duscher[1], Sergei Kalinin[1*]

[1] Department of Materials Science and Engineering, University of Tennessee, Knoxville, TN 37923, USA
[2] Department of Chemistry, University of Washington, Seattle, WA 98195, USA
[3] Exabyte Inc. (Mat3ra.com), Walnut Creek, 94596 CA, USA
[4] Johns Hopkins University, Baltimore, MD, 21218, USA

The real-world implementation of materials prediction algorithms remains limited by persistent characterization bottlenecks in materials discovery, where photon-based probe techniques (e.g., XRD or Raman) impose long acquisition times and access latencies, restricting exploration to quasi-ternary composition spaces typically realized as compositional libraries. Here, we argue that a paradigm shift from photon- to electron-based characterization can realign materials characterization with modern high-throughput synthesis. We formulate cost functions and exploration strategies for STEM-based chemical and structural characterization and use Monte Carlo simulations to show that random chemical libraries, where compositionally distinct regions are co-located within a single specimen and interrogated in situ by electron spectroscopies, can sample high-dimensional composition and phase spaces with orders-of-magnitude greater effective coverage than conventional spread-library/X-ray approaches. We further demonstrate autonomous discovery on a laboratory STEM platform, where ML-based autotuning and scripted control enable iterative region selection and characterization without human intervention. Finally, we outline extensions to labeled or position-encoded libraries that preserve compositional and processing metadata, enabling joint exploration of composition and process spaces. Together, these results establish electron-based, ML-enabled STEM as a scalable pathway toward combinatorially rich materials discovery.

## I. Introduction

Accelerated materials exploration and deployment in industry is widely recognized as a major priority for the coming decade. The Materials Genome Initiative, launched in 2011, catalyzed more than a decade of progress in theory and simulation, with large-scale computational workflows and data infrastructures being adopted across academia and industry.[1-4] In the last five years, major technology companies such as Google, Microsoft, and Meta have entered the materials prediction space, greatly expanding the reach of data-driven and physics-informed models.[5-7] Yet it has become increasingly clear that real materials defined by defects, microstructures, and manufacturing histories, and their functional responses often emerge at length scales above the atomic and must ultimately be validated and realized experimentally. Over the last two to three years, this realization has driven a rapid pivot from "prediction only" paradigms toward experimental materials exploration at scale, with efforts such as Google and Amazon's experimental materials discovery labs and substantial venture investments (e.g., Lila Scientific, Periodic Labs) signaling a renewed focus on building physical discovery pipelines.[8, 9]

The scale and urgency of these investments, however, obscure the fact that this is not the first wave of enthusiasm for accelerated experimental discovery. Much like the cycles of optimism and retrenchment in artificial intelligence, accelerated materials exploration has experienced its own "winters." The first wave in the 1960s–1970s involved diffusion couples and diffusion junctions as platforms for probing phase evolution in alloys along continuous compositional gradients.[10, 11] A second wave emerged in the 1990s and early 2000s, when advances in physical deposition methods such as pulsed laser deposition (PLD) and sputtering enabled combinatorial spread libraries that sampled binary and ternary sections of high-dimensional phase diagrams.[12-15] Despite impressive early demonstrations, this combinatorial effort was followed by nearly two decades of limited support, with only a small number of groups maintaining comprehensive experimental programs.

The last 6 years have marked a qualitatively different phase, characterized not only by renewed enthusiasm but by the practical commodification of high-throughput and ML-controlled synthesis.[16-20] Whereas before 2020 demonstrations typically depended on in-house engineered microfluidic or feedback-controlled platforms – such as automated perovskite nanocrystal screening[21] or continuous-flow reaction optimization systems[22, 23] that required substantial custom mechanical, electronic, and software integration – post-2020 efforts increasingly leverage modular, commercially available robotics and orchestration software. Low-cost liquid handlers, configurable chemical automation workstations, and collaborative robotic handlers now allow academic laboratories to execute plate- or vial-based libraries, integrate routine analytics, and implement closed-loop Bayesian or active-learning control largely with off-the-shelf components.[18, 20] In parallel, ML-driven optimization has moved from proof-of-concept to reproducible performance gains, including Bayessian Optimization (BO) surpassing expert-guided reaction design,[24] autonomous batch and flow synthesis platforms,[25] reinforcement-learning-controlled multi-step chemistry,[26] and autonomous solid-state synthesis systems achieving high-throughput discovery rates.[27, 28]

With the commodification of accelerated synthesis, the dominant bottleneck has shifted decisively to characterization. For any new material, structural characterization, traditionally based on photon-driven probes such as X-ray diffraction and optical spectroscopy, is the accepted first step, followed by functional measurements (mechanical, ferroelectric, electrochemical, photovoltaic, etc.) that define exploration and application domains. However, many of these measurements are intrinsically slow and constrained by the physics of the probe and the design of the instrumentation: laboratory X-ray diffraction, the cornerstone of structure determination, often requires 10–30 minutes per spectrum; more complex property measurements, such as mechanical, ferroelectric, or electrochemical, take longer. As long as our primary characterization channels are photon-based techniques with relatively low information density per unit time, advances in synthesis throughput will not translate directly into accelerated discovery.

This motivates a transition in the experimental pathway to discovery from photons to electrons, leveraging electron- and electron-based spectroscopic probes that can access microstructure, defects, and local functionality at the length scales where real materials behavior emerges, and that can, in principle, be embedded in automated, high-throughput, and information-rich workflows. However, electron microscopies are traditionally believed to be very low-throughput tools, both due to sample preparation logistics and to the time-consuming, manual instrument tuning and optimization. Over the last 5 years, rapid progress in automated electron microscopy image optimization and image space exploration illustrates that the second constraint is well on its way to being addressed by ML-enabled microscope control.[29, 30]

Here, we pose that the multimodal imaging nature of STEM, which allows simultaneous chemical and structural imaging, opens the pathway for massive multiplexing of chemical spaces that can be explored within STEM, naturally building upon the high-throughput capabilities enabled by lab robotics.

## II. Strategies for materials space exploration

It is essential to design theory–experiment workflows that are explicitly cognizant of latency, cost, and accessibility across chemical and synthesis spaces. Theory-driven materials exploration offers a fundamentally different scaling behavior than experiment: in principle, any region of compositional or chemical space can be accessed with a relatively uniform "transition cost," with computational expense determined primarily by system size, model fidelity, and available computing resources rather than by the specific chemistry under study. High-throughput first-principles calculations have been demonstrated for hundreds of binary compounds with standardized, accessible workflows that make data repeatable and extensible for downstream modeling and analysis.[31] Beyond raw throughput, integrating machine learning with physical models highlights the importance of model *interpretability* when connecting structure and properties, as shown by recent collaborative work emphasizing interpretable machine learning for materials design.[32] Similarly, structured, machine-readable representations of computational models and results are key for building interoperable data ecosystems that support large-scale theory workflows and subsequent integration with experiment.[33] However, this apparent uniformity of theoretical access increasingly contrasts with experimental reality, where synthesis constraints, processing histories, and characterization latencies impose highly nonuniform costs. Bridging this gap requires experimental platforms whose throughput, data structure, and accessibility more closely mirror the effective scalability of modern theory, enabling tightly coupled, latency-aware closed-loop discovery workflows.

By contrast, experimental synthesis is constrained by the specifics of the platform and the system under study. In molecular discovery, navigation of chemical space can be framed as movement on a graph of known compounds and reactions, with retrosynthesis (classical and ML-based) charting viable routes to targets and reaction optimization framed as a relatively low-dimensional control problem.[34, 35] Within this setting, prediction-to-realization pipelines have already achieved notable successes, as evidenced by multiple industrial efforts in molecular design and drug discovery.[36, 37]

Functional inorganic materials pose a different challenge, both in realizability and in control, because microstructure from the atomic scale to the mesoscale and beyond becomes a defining and often dominant factor. Synthesis routes and micro- nanostructures are co-designed: processing conditions (e.g., temperature, atmosphere, strain, cooling rate) determine phase assemblages, grain structures, domain configurations, and defect populations. Classic examples include steels and structural alloys, where centuries of practice have effectively optimized the microstructure for performance, as well as relaxor ferroelectrics, oxide ceramics, and many functional oxides whose behavior is governed by intricate structural motifs on a small length scale. In these systems, composition alone is not sufficient; the path to discovery requires resolving and controlling the underlying micro- and nanostructure.

At the same time, materials synthesis itself is subject to stringent constraints on scale-up and flexibility. Solution-based robotics can increase throughput by factors of $10^2$–$10^3$, but typically within a fixed set of endmembers; incorporating new chemistries often demands substantial re-engineering. Automated ceramic synthesis offers perhaps an order-of-magnitude improvement over manual workflows but remains limited by the intrinsic complexity of powder processing and thermal treatments. Combinatorial spread libraries provide efficient sampling of quasiternary two-dimensional manifolds in high-dimensional spaces, yet they cannot fully span the broader design space, and many processing steps are difficult or impossible to multiplex beyond gradient furnaces or localized treatments such as laser annealing. These platform-specific constraints impose strong, domain-dependent limitations on how experimental campaigns for discovery and optimization can be planned.

## III. STEM for random library exploration

Scanning transmission electron microscopy (STEM) provides a uniquely rich, electron-based window into materials behavior at the atomic scale.[38-40] In a single platform, it can probe atomic structure and defects via high-angle annular dark-field (HAADF)[41, 42] and phase-contrast imaging,[43] leading strain mapping[44] and structure mapping[45] with picometer-precision. With modern detectors, it is common to map chemical composition using energy-dispersive X-ray spectroscopy and electron energy loss spectroscopy, and access vibrational excitations and even local temperature via low-loss electron energy-loss spectroscopy (EELS).[46] Core-loss EELS further resolves oxidation states,[47] coordination environments,[48, 49] and bonding anisotropy,[50] while advanced modalities such as 4D-STEM[51] capture local strain,[52] electric and magnetic fields,[53, 54] and scattering signatures of order–disorder transitions.[55, 56] Combined with powerful computational methods, the electron microscope can resolve atoms down to their vibrational limit via ptychography[57] and reconstruct the atomic configuration of nanoparticles via tomography.[58] Historically, exploiting this information density required painstaking, manual optimization of alignment, aberration correction,[59, 60] and detection conditions for each experiment.[61] Recently, however, the community has been pushing towards machine-learning–enabled autotuning and adaptive control to automate these steps,[62] dramatically reducing setup times and opening the door to high-throughput, STEM-based discovery workflows.[30, 63-65]

Despite its exceptional information content, STEM remains fundamentally limited by sample throughput. Preparing suitable thin, clean, and mechanically stable specimens is labor-intensive and often sample-specific, involving focused ion beam (FIB) milling, polishing, or delicate exfoliation and transfer steps.[66, 67] Once in the microscope, each region of interest must be located, aligned, and imaged or spectroscopically mapped under carefully tuned conditions, with drift correction and dose management further constraining acquisition speed. As a result, even in well-optimized workflows, only a handful of distinct samples or regions can typically be examined in detail within a day.

An alternative route to increasing the effective reach of STEM is to move from carefully designed single-composition specimens to *random libraries*. In this approach, multiple compositions, phases, or processing conditions are pooled into a single heterogeneous specimen, for example, by co-depositing or physically mixing powders, fragments, or particles obtained from different synthesis batches. In many practical implementations this naturally leads to nanoparticle or nanostructured mixtures, allowing multiple materials to be probed within a single STEM field of view. Rather than knowing *a priori* which region corresponds to which material, we rely on STEM's chemical sensitivity via EDS or EELS to identify local composition and phase at each probed region. Once individual phases or compositional domains are recognized in this way, their atomic structure, defect landscapes, and vibrational properties can be interrogated within the same dataset. Random libraries thus trade spatial order for informational density, leveraging STEM's capability to chemically fingerprint nanoscale regions in order to survey many materials and microstructures in a single experiment.

The key expected advantage of the random-library approach is the expansion of the accessible compositional space encoded within a single specimen. However, practical STEM implementations impose constraints on sample size, which is typically limited by the supporting substrate, such as a TEM grid. As a result, the accessible spatial area for STEM characterization is usually restricted to a few square millimeters. Consequently, a first essential step is to estimate the effective dimensionality of the compositional space that can be represented within such a random library under realistic sample-size constraints. Importantly, sample preparation is a complex, multivariate process that depends on numerous experimental conditions and material properties and is rarely captured by simple mathematical models; therefore, the estimates presented below should be interpreted as conservative, order-of-magnitude bounds rather than precise predictions.

For this simplified model, we adopt the following assumptions: (i) particles do not overlap, (ii) particles are uniformly distributed over the surface, (iii) particle sizes follow a Gaussian distribution with a standard deviation of 30% of the mean size, and (iv) particle shape effects are neglected by characterizing particles using an effective radius, such that particles are treated as spherical (or circular in the two-dimensional representation). To define a conservative lower-bound scenario, we impose a 50% area-coverage threshold, assuming that particles occupy 50% of the available sample surface while the remaining 50% is free space; this coverage is approximately 40% below the theoretical close-packing limit for circular particles. Compositional sampling is performed on a compositional simplex using a uniform grid with a step size of 0.1, corresponding to a minimum of 10 sampling points along each compositional dimension. Assuming a 2 × 2 $mm^2$ TEM grid and taking only two-thirds of the grid as usable (with the remainder occupied by the frame), the available area for particle deposition is approximately 2.67 $mm^2$.

Under these assumptions, the expected number of particles, and thus an upper bound on the dimensionality of the compositional space representable by a random particle library deposited on a TEM grid, can be estimated as

$$\mathbb{E}[N(\mu)] = \frac{A_c}{\pi(\mu^2+\sigma^2)}, \quad (1)$$

where $A_c$ is the particle-covered area of the grid, $\mu$ is the mean particle radius, and $\sigma$ is the standard deviation of the particle radius distribution. From this scaling, we find that effective particle radii of approximately 500 nm, 150 nm, and 50 nm are sufficient to represent 6D, 7D, and 8D compositional spaces, respectively (Figure 1). Although these values represent optimistic limits that may not be fully attainable in practice, they nevertheless demonstrate that random-library samples can, in principle, enable exploration of compositional spaces at least three dimensions higher than those accessible with conventional spread-library approaches.

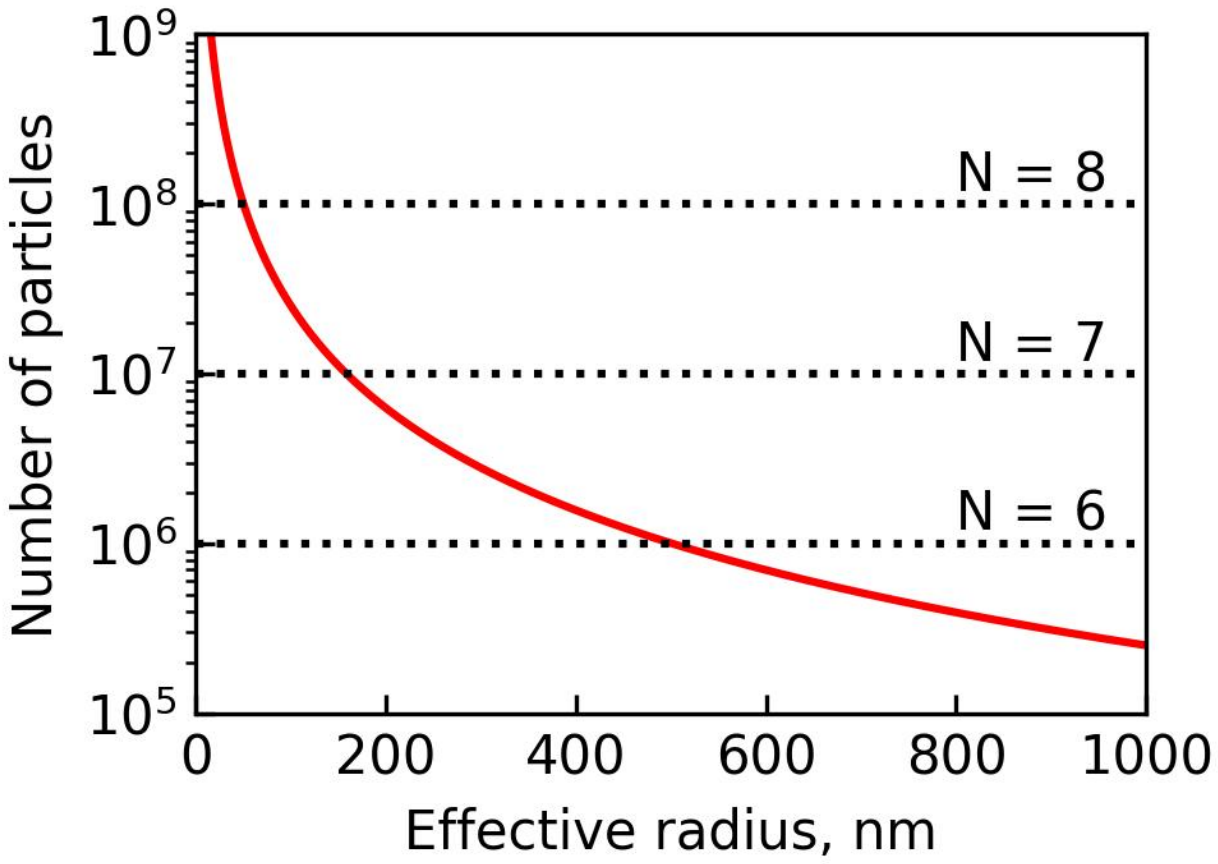

**Figure 1.** Expected particle count as a function of mean their radius for the random libraries. Particle sizes are assumed to follow a Gaussian distribution with a standard deviation of 30% of the mean radius ($\sigma = 0.3\mu$). Full area $A = 2.67$, area-coverage – 0.5.

**IV. Realistic optimization workflow for random libraries**

Here, we introduce a realistic workflow for STEM-based compositional optimization and discovery in a random particle library (Figure 2). We consider autonomous exploration of a random library, in which each particle is characterized by a composition vector **c** sampled from an $N$-dimensional compositional space. A representative model system is a multicomponent alloy (e.g., $A_xB_yC_{1-x-y}$ or $A_xB_yC_zD_{1-x-y-z}$), subject to the constraint that the elemental fractions sum to unity. Compositional characterization is performed using energy-dispersive X-ray spectroscopy (EDX). For functional characterization, we consider scanning transmission electron microscopy electron energy-loss spectroscopy (STEM-EELS) as a model technique for estimating the target response of individual particles. The EELS workflow is treated as a serial process in which, at each iteration, the electron beam is positioned on a selected particle to acquire a spectrum from which the target functionality is extracted. EELS acquisition times can vary widely depending on the spectral range, signal-to-noise requirements, and the physical properties of interest. Same logic can be applied to the local structure determination via diffraction, or any of the advanced spectroscopic modalities available in STEM.

Experimentally, the exploration begins with an overview scan to locate and identify particles within the FoV. This scan provides particle positions and approximate sizes but does not yield compositional information. To determine particle compositions, EDX mapping is subsequently performed within the same FoV. The resulting elemental fractions $\{c_i\}_{i=1}^{N}$ are associated with element-specific uncertainties $\{\sigma_i\}$, reflecting detector sensitivity and counting statistics. The measured compositions are constrained by normalization, $\sum_{i=1}^{N} c_i = 1$. To exploit this constraint, we identify the component $k$ with the largest measurement uncertainty and re-estimate its concentration from the remaining components as

$$c_k = 1 - \sum_{i \neq k} c_i. \quad (2)$$

Assuming independent uncertainties for the remaining $N - 1$ components, the uncertainty of the re-estimated component follows from standard error propagation:

$$\sigma_k = \left( \sum_{i \neq k} \sigma_i^2 \right)^{1/2}. \quad (3)$$

We define a scalar measure of the total compositional uncertainty for each particle as the root-sum-square of the uncertainties of the independently measured components, which is numerically equal to the uncertainty of the re-estimated component $\sigma_{tot} = \sigma_k$. This procedure enforces the simplex constraint, avoids double-counting poorly determined components, and yields a more stable and physically consistent estimate of compositional uncertainty for subsequent analysis.

Given the approximate particle compositions and spatial locations identified within the accessible field, EELS-based functional exploration is carried out using cost-aware Bayesian optimization (BO) in composition space, where the set of particles discovered within a limited spatial region defines the accessible candidate compositions. At any given time, BO therefore operates only on this locally available subset of the combinatorial library. When candidate particles lie beyond the beam-shift–accessible range, the additional cost associated with motorized stage motion is explicitly accounted for using the motion cost function that will be introduced in Section IV.1. Accordingly, the selection of the next particle to be measured is governed not only by the composition-based BO acquisition function but also by the spatial cost of repositioning in real space. To capture this trade-off, we define a cost-aware acquisition score

$$S(c_i) = \frac{a_t(c_i)}{C_{\text{move}}(\mathbf{r}_i)}, \quad (4)$$

where $a_t(c_i)$ is the BO acquisition value for the composition $c_i$, and $C_{\text{move}}(\mathbf{r}_i)$is the motion cost associated with moving the probe to the spatial coordinates $\mathbf{r}_i$of the corresponding particle. This formulation enables simultaneous optimization in composition space and real space by prioritizing measurements that offer high expected information gain relative to their experimental cost.

However, the particles contained within a single accessible region represent only a small fraction of the full combinatorial library distributed across the substrate. Consequently, efficient optimization requires the ability to adaptively transition between spatial regions in order to discover and characterize new particles. To account for both the potential compositional gain enabled by spatial exploration and the non-negligible cost associated with global stage motion, we need to introduce an additional metric that quantifies the expected benefit of switching to a new spatial region after each BO iteration, i.e., after each particle measurement. While BO governs the selection of the next particle within the currently accessible region based on composition-space utility, region switching is treated as a higher-level control decision that balances continued local exploitation against the expected information gain from accessing previously unexplored areas of the sample. In the following Section IV.2, we describe the decision logic for remaining in the current region versus moving to a new spatial region.

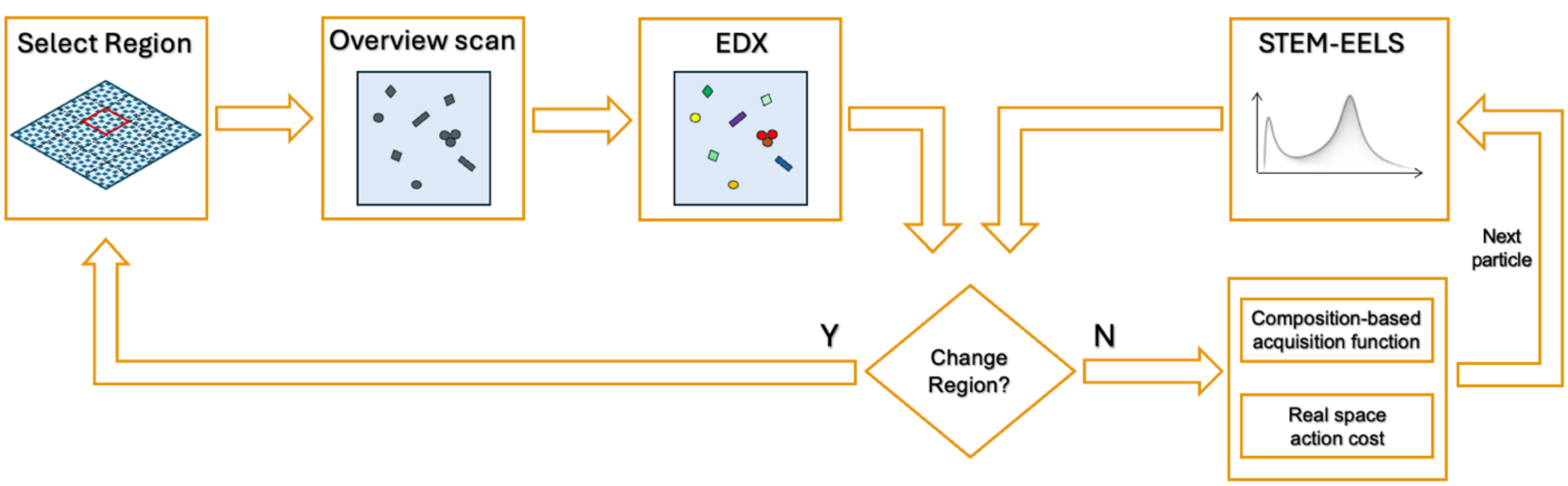

**Figure 2.** Random library characterization workflow

### IV.1. Cost Function Design

The transition from purely computational and data-driven workflows to real-world instrumentation requires explicit consideration of a cost function. This cost function quantifies the resources required to execute available agent actions, such as spectral acquisition, positioning, and instrumental tuning, in terms of time or other investments. Incorporating such a cost function enables balancing the effort associated with different actions against their expected informational or scientific return.

Consequently, assessing the scalability of the random-library approach for high-throughput materials design requires careful estimation of instrumental limits and latencies, which can be explicitly incorporated through the cost function. Along similar lines, Chawla et al. analyzed cost-aware experimental design in nanoindentation, assuming that local chemical composition was known from prior fast energy-dispersive measurements and that measurement cost was governed by instrumental constraints.[68] Slautin et al. introduced time-aware Bayesian optimization for spectroscopic techniques, explicitly accounting for the trade-off between acquisition time and resulting measurement precision.[69]

Introducing a cost function for STEM-based random-library exploration requires explicit consideration of the time scales associated with different microscope operations and measurement modes. Depending on the operating mode, STEM can function either in an imaging regime, where the electron beam repeatedly scans a defined field of view, or in a spectroscopic regime, where spectra are acquired from a selected region over a specified dwell time. Imaging modes, which in our case can be used for particle localization or coarse compositional mapping (e.g., via EDX), provide spatially distributed information across the entire field of view, capturing all particles within the region. In contrast, spectroscopic techniques typically yield higher-fidelity chemical or electronic information but are limited to single locations. These modes are therefore complementary rather than interchangeable, and direct comparison of their costs is not meaningful. Instead, imaging-based particle localization and compositional profiling are often prerequisite steps for subsequent functional characterization using spectroscopic techniques such as EELS. As a result, any algorithm for autonomous functional characterization must ultimately account for the cumulative cost of particle localization, compositional profiling, and targeted spectroscopic acquisition.

The physical extent of a random library may exceed the available field of view (FoV), requiring explicit exploration via motorized stage translation for global positioning. Within a given FoV, the time required to reposition the electron beam between particles is dominated by electric and magnetic field reconfiguration and is negligible compared to other latencies in the experimental workflow. Consequently, beam repositioning time is not treated as a limiting factor in our model. Instead, the dominant intra-FoV latency arises from information exchange and control overhead, including data transfer, processing, and communication with the control server, which we assume to be on the order of 10–30 ms per action. The beam shift-accessible area can exceed the nominal FoV by a factor of several. This capability enables more complex workflows in which particle mapping and compositional characterization are performed over extended regions.

In contrast, motion beyond the beam-shift range requires motorized stage translation and introduces substantial additional delays. Given an average stage speed $v$, the corresponding motion time scales linearly with the displacement distance $d$. We model the total repositioning time from the center of the current FoV as

$$t_{\mathrm{move}}(d) = \begin{cases} t_{\mathrm{bs}}, & d \le d_{\mathrm{th}}, \\ t_0 + d/v, & d > d_{\mathrm{th}}, \end{cases} \tag{5}$$

where $t_{\mathrm{bs}}$ denotes the effective beam-shift repositioning time within the accessible range $d_{\mathrm{th}}$, and $t_0 = t_{\mathrm{ma}} + t_{\mathrm{bs}}$ accounts for fixed latencies associated with global stage motion, including microscope realignment, overview image acquisition for particle localization, and, in some cases, compositional measurements.

Based on this timing model, we define a motion cost function for functional characterization of random particle libraries as

$$C_{\mathrm{move}}(d) = \begin{cases} c_{\mathrm{intro}}, & d \le d_{\mathrm{th}}, \\ \alpha \left( t_0 + \frac{d}{v} \right), & d > d_{\mathrm{th}}, \end{cases} \tag{6}$$

where $\alpha$ is a linear scaling coefficient that converts physical time into an effective exploration cost, and $c_{\mathrm{intro}}$ represents the cost of an intra-FoV position change, which we set to unity. This formulation captures the sharp increase in cost associated with leaving the current FoV and provides a physically motivated mechanism for balancing local exploitation against global spatial exploration.

**IV.2. Cost-Aware Spatial Region Switching**

This region-switching metric explicitly balances two competing factors: (1) the *remaining opportunity within the current field of view* (ROV), which decreases as particles are progressively measured, and (2) the *expected gain from exploring a new region*, which depends on the likelihood of encountering particles with compositions that offer high acquisition value under the current BO model. Switching to a new region incurs additional experimental cost associated with motorized stage motion, realignment, and EDX particle composition discovery. By evaluating this region-level advantage after each measurement, the framework enables adaptive control of the accessible field, ensuring that stage motion is initiated only when the expected benefit of exploring a new region outweighs both the remaining local opportunity and the associated experimental overhead.

The remaining opportunity within the current field of view is quantified by the ROV, which estimates the expected benefit of continuing measurements in the presently accessible region. Let $N_{\text{tot}}$ denote the total number of particles identified within the current region, and $N_{\text{meas}}$ the number of particles already characterized by EELS. The fraction of remaining, unmeasured opportunity is then

$$R = 1 - \frac{N_{\text{meas}}}{N_{\text{tot}}}. \tag{7}$$

At a given iteration $i$, the BO surrogate model defines an acquisition function $a_i(c)$ over composition space. Restricting attention to the set of unmeasured particles within the field of view, the maximum attainable acquisition value is

$$G(i) = \max_{i \in \mathcal{U}_{\text{unm}}} a_t(c_i), \tag{8}$$

where $\mathcal{U}_{\text{unm}}$ denotes the set of unmeasured particles and $c_i$ their compositions. The ROV is then defined as

$$V_{\text{stay}}(i) = R \cdot G(i) \tag{9}$$

and represents the expected utility of continuing exploration within the current field of view. This formulation naturally accounts for particle depletion: as measurements accumulate, $R_{\text{FoV}}$ decreases, reducing $V_{\text{stay}}$ even if the instantaneous acquisition landscape remains favorable. Consequently, $V_{\text{stay}}$ provides a principled, dynamically updated measure of diminishing local opportunity that can be directly compared to the expected value of switching to a new spatial region.

The expected benefit of switching to a new spatial region is quantified by the Expected Switch Gain, denoted $E_{\text{gain}}$, which captures the potential utility of accessing particles that are not yet observed. Prior to moving the stage, neither the number of particles nor their compositions in an unexplored region are known. We therefore model the particle discovery process statistically. For a candidate region of area $A$, the number of particles $N$ is modeled as a Poisson random variable, assuming independent particle occurrences uniformly distributed in space:

$$N \sim \text{Poisson}(\rho A), \tag{11}$$

where $\rho$ is the particle density, defined by the already explored region. Conditional on $N$, particle compositions $\{c_i\}_{i=1}^{N}$ are sampled from a prior compositional distribution $p(c)$ reflecting the synthesis variability of the library. Given the current BO acquisition function $a_i(c)$, the potential gain associated with the new region is defined as the maximum acquisition value among the newly discovered particles,

$$G_{\text{new}}(i) = \max_{j=1,\ldots,N} a_i(c_j). \tag{12}$$

The expected switch gain is then obtained by averaging over the uncertainty in both particle count and composition,

$$E_{\text{gain}}(i) = \mathbb{E}[G_{\text{new}}(i)], \tag{13}$$

which is estimated numerically via Monte Carlo sampling.

In addition to acquisition-driven utility, the expected gain from switching regions may incorporate a compositional novelty component, which explicitly favors exploration of previously unobserved areas of composition space. This term reflects the fact that, as the experiment progresses, the marginal value of discovering compositions similar to those already measured decreases, even if their instantaneous acquisition value remains moderate. To capture this effect, we introduce a novelty score $n(c) \in [0,1]$ for a candidate composition $c$, defined as a monotonically increasing function of its distance from the set of previously measured compositions $C_{\text{seen}}$,

$$n(c) = 1 - \exp\left(-\frac{d(c, C_{\text{seen}})}{d_0}\right), \tag{14}$$

where $d(c, C_{\text{seen}})$ denotes the minimum distance in composition space and $d_0$ is a characteristic length scale, taken equal to maximum possible distance within the motorized stage accessible range. The acquisition function is then augmented as

$$\tilde{a}_i(c) = a_i(c)\,(1 + \lambda_{\text{nov}}\, n(c)), \tag{15}$$

where $\lambda_{\text{nov}}$controls the relative weight of novelty. The expected gain of switching to a new region is computed using $\tilde{a}_i(c)$in place of $a_i(c)$.

Switching to a new region incurs an experimental time cost associated with motorized stage motion, microscope re-alignment, and renewed particle discovery and compositional characterization. Let $t_{\text{switch}}$ denote the total time required for this transition. During this interval, functional measurements cannot be performed, resulting in an opportunity cost proportional to the current local gain rate. This cost is expressed as

$$C_{\text{switch}}(t) = t_{\text{switch}} \cdot \frac{V_{\text{stay}}(t)}{t_{\text{meas}}}, \tag{16}$$

where $t_{\text{meas}}$ is the time required for a single EELS measurement. The net value of switching to a new region is then defined as

$$V_{\text{switch}}(t) = E_{\text{gain}}(t) - C_{\text{switch}}(t), \quad (17)$$

which directly balances the expected benefit of discovering new particles against the experimental overhead associated with stage motion. This formulation enables a quantitative comparison between continuing local exploration and initiating a region switch at each iteration. A region switch is initiated whenever the expected net benefit of switching is positive, $V_{\text{switch}}(t) > 0$; otherwise, exploration continues within the current FoV.

### IV.3. Simulations

To demonstrate the proposed STEM-based optimization workflow for random libraries, we performed a series of Monte Carlo simulations under realistic experimental constraints representative of modern STEM instrumentation. The model system was defined as a multidimensional alloy with uniform sampling over the compositional space, where particles were represented as spherical with an average effective radius of 20 nm and a standard deviation of 10 nm. To limit computational complexity while preserving realistic spatial structure, the globally accessible area was restricted to $300 \times 300$ μm$^2$, within which particles were uniformly distributed with density ~0.02 particle/μm$^2$.

Compositional characterization was assumed to be performed via EDX mapping, with a total acquisition time of 10 s per frame, representing a practical balance between throughput and compositional precision. Functional characterization was modeled using STEM-EELS, with a fixed acquisition time of 1 s per particle, consistent with measurements targeting electronic or near-edge spectral features. The beam-shift accessible area has been limited by the FoV $30 \times 30$ μm$^2$. Global stage motion was assumed to occur at a speed of 10 μm/s, and a retuning time of 5 s was included following each change of the local measurement region. The overall measurement cost was computed using the cost function defined in Equation 6, with $\alpha = 1$.

The target functionality probed by STEM-EELS was modeled as a superposition of 106 Gaussian functions with a standard deviation of 0.1 defined over the compositional space, generating a complex, multimodal response landscape. The autonomous experiment was guided by a pure-exploration strategy using a cost-weighted maximum-uncertainty acquisition function.

The optimization was initialized using three randomly selected seed particles, followed by Bayesian optimization iterations.

In the first set of simulations, we model STEM-based exploration of a three-dimensional compositional space. As a representative system, we consider a model Au-Co-Ni alloy (Figure 3). The specific elemental choice only determines the achievable precision of compositional estimation and does not affect the generality of the workflow. The target functionality is defined using the synthetic response function described above. Following the initial seed measurements, the optimization proceeds for a total of 150 BO iterations.

The experimental trajectory in compositional space exhibits an approximately uniform distribution of sampled points across the compositional simplex, indicating efficient exploration of the accessible composition space (Figure 3b). Consistent with this behavior, both the mean predictive uncertainty and the absolute error with respect to the ground-truth target function decrease monotonically as the experiment progresses, reflecting the expected convergence characteristics of an exploration-driven BO strategy (Figure 3c,d).

More informative trends emerge when examining the experimental trajectory in real space (Figure 3a). The spatial trajectory clearly follows the anticipated logic of the workflow: repeated exploration of particles within the beam-shift–accessible FoV, followed by discrete shifts of the FoV after a certain number of iterations. The fraction of particles explored within each FoV decreases over time and saturates at approximately 50–60% (Figure 3e). This trend can be attributed to the increasing selectivity of the surrogate model as knowledge accumulates: during the early stages of the experiment, limited prior information renders most particles within the FoV informative, whereas continued exploration progressively reduces the unexplored compositional space, leading the algorithm to select an increasingly smaller subset of candidate particles at each iteration.

During transitions between FoVs, the algorithm preferentially selects neighboring spatial regions for stage movement. This behavior arises from the assumed uniform particle density across the sample, which renders the expected informational value of different regions comparable, leaving the stage-motion cost, which is proportional to travel distance, as the dominant factor. Under alternative assumptions, such as partial awareness of global compositional trends or spatially non-uniform particle densities, the algorithm would naturally favor more distant regions offering higher expected informational gain despite the increased motion cost.

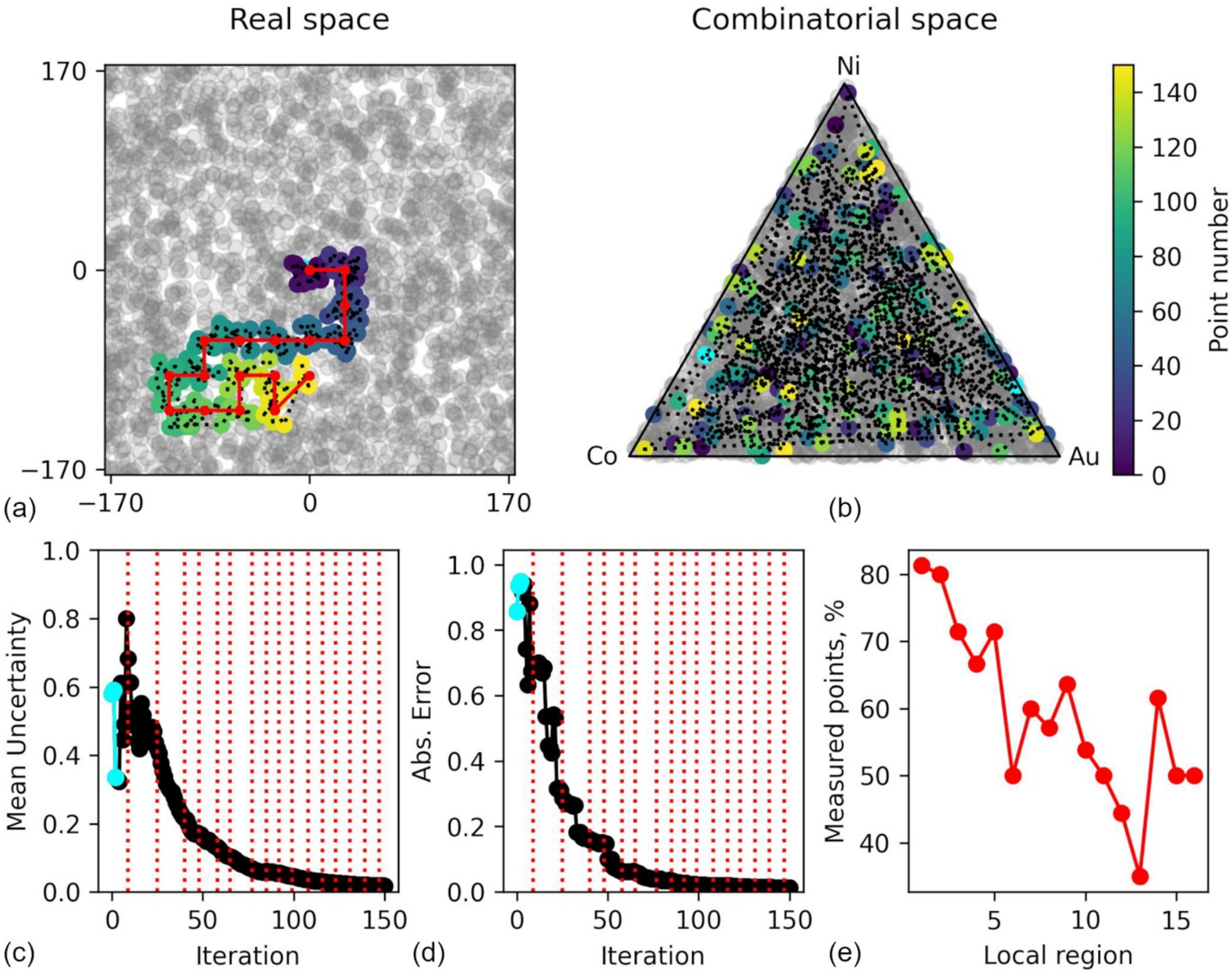

**Figure 3.** Simulation of STEM-based exploration of a random library in a three-dimensional compositional space. (a) Experimental trajectory in real space. (b) Experimental trajectory in compositional space. (c) Evolution of the mean predictive uncertainty. (d) Absolute error with respect to the ground-truth target function during the experiment. (e) Fraction of particles measured within each local region. The solid red line in (a) indicates the motorized stage positioning trajectory. Dashed vertical lines in (c,d) mark iterations at which the local region was changed. The experiment was initialized from three seed points, indicated by cyan markers.

To demonstrate the generality of the proposed approach, we extended the framework to four- and five-dimensional compositional spaces (Figure 4). Increasing the dimensionality leads to an exponential growth of the accessible compositional space, necessitating longer optimization trajectories. Accordingly, the number of BO iterations was set to 300 for both cases. Visualization of trajectories in high-dimensional compositional spaces becomes impractical. The analysis was restricted to the experimental trajectories in real space and the primary evolutionary metrics.

Both the four- and five-dimensional simulations exhibit trends consistent with those observed in the three-dimensional case, including a gradual decrease in mean predictive uncertainty (Figure 4b,f), absolute error (Figure 4c,g) with respect to the ground-truth target, and the fraction of measured points over time. Notably, in both higher-dimensional cases, the fraction of measured particles continues to decrease steadily without reaching saturation within the explored iteration range. Moreover, the reduction in the fraction of measured particles per spatial region occurs more slowly as the compositional dimensionality increases, which can be attributed to the slower growth of model selectivity in larger, higher-dimensional compositional spaces.

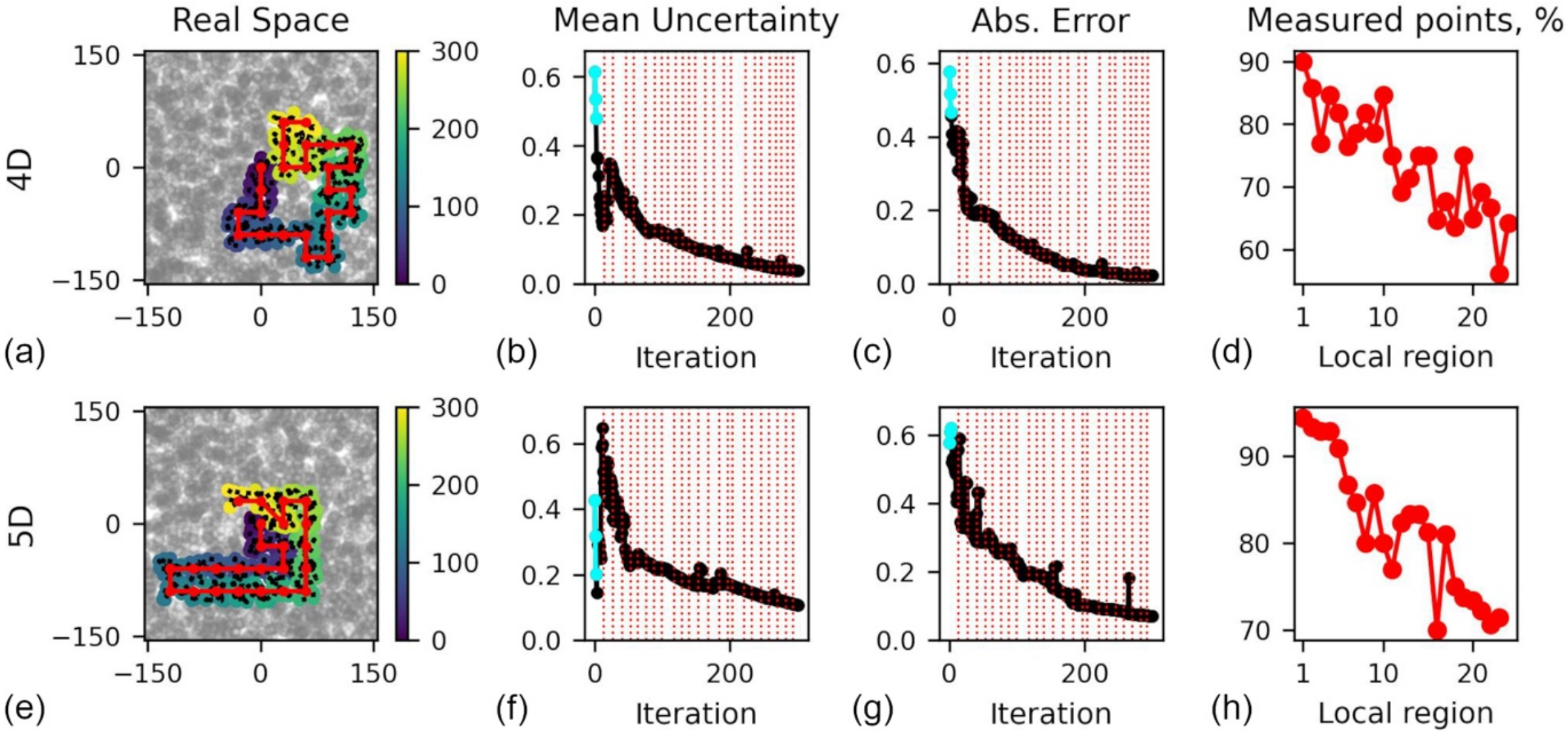

**Figure 4.** Simulation of STEM-based exploration of a random library in (a–d) four-dimensional and (e–h) five-dimensional compositional spaces. (a, e) Experimental trajectories in real space. (b, f) Evolution of the mean predictive uncertainty. (c, g) Absolute error with respect to the ground-truth target function during the experiments. (d, h) Fraction of particles measured within each local region. The solid red line in (a, e) indicates the motorized stage positioning trajectory. Dashed vertical lines in (b, c, f, g) mark iterations at which the local region was changed. The experiments were initialized from three seed points, indicated by cyan markers.

### IV.4. Perspective

Overall, the Monte Carlo simulations demonstrate the applicability and scalability of the proposed framework for efficient exploration of high-dimensional compositional spaces under realistic experimental constraints. Across increasing dimensionalities, the framework consistently balances local and global spatial exploration, yielding systematic reductions in predictive uncertainty and error while maintaining manageable measurement costs. These results indicate that cost-aware, spatially adaptive STEM workflows can sustain effective discovery even as the dimensionality of the compositional space grows.

In this setting, it is important to recognize that the relationship between initial mixture composition and the phases that actually form can be highly nontrivial, governed by growth kinetics, segregation, and metastability; STEM "reads out" what phases and microstructures are present locally, but not directly what was originally put into the system. This ambiguity is not unique to random libraries and is, in fact, shared with conventional combinatorial approaches, where diffusion, reactions, and processing histories can blur the link between nominal and realized compositions. Spread libraries partially mitigate this issue through positional encoding: the spatial coordinates on the substrate provide a prior on composition and processing conditions. Looking forward, one can envision hybrid strategies in which random or partially ordered libraries are augmented by encoded particles or chemical or structural markers that preserve information about synthesis or processing history, thereby enabling similar high-throughput approaches to be extended into processing spaces as well as composition spaces.

We also emphasize that many parameters in the Monte Carlo model are either deliberately simplified or known only within limited accuracy. Process temperatures, diffusion barriers, interaction energies, and reaction probabilities are typically treated in coarse-grained or phenomenological form, reflecting incomplete microscopic knowledge. Rather than viewing these approximations as fixed limitations, we regard them as opportunities for learning from experiment: by comparing simulated and measured structural, chemical, or spectroscopic observables across the random library, the MC parameters can be systematically inferred or refined. In this way, random-library STEM measurements and MC modeling form a closed loop, in which experiments not only validate predictions but also calibrate and improve the underlying stochastic model.

## V. Experimental feasibility

### V.1. Model system

To demonstrate the experimental feasibility of the proposed random-library STEM workflow, we employ a heterogeneous nanoparticle ensemble composed of chemically and structurally distinct nanostructures, including CdSe and CdS quantum dots, CdSe nanoplatelets, CdS nanorods, and InP-based nanoparticles. This deliberately mixed system serves as a simplified model random library in which multiple compositions and morphologies coexist within a single specimen. The objective of this demonstration is to validate the practical applicability of the workflow rather than to realize its full closed-loop implementation. Specifically, we focus on the foundational steps required for autonomous exploration: particle identification and reliable compositional determination, while leaving fully integrated cost-aware optimization and region-switching strategies for future work.

We begin the experimental realization with particle identification. As a proof-of-principle example, we employed the Segment Anything Model (SAM) for rapid particle segmentation on HAADF-STEM images (Figure 5).[70] Importantly, this was a dummy implementation: we used the out-of-the-box SAM model without extensive hyperparameter tuning, domain-specific retraining, or post-processing refinement. No advanced filtering, morphology correction, or drift compensation was applied. The purpose here is only to evaluate whether particle identification constitutes a practical bottleneck for random-library STEM exploration.

As shown in Figure 5, the majority of nanoparticles are clearly distinguished and segmented. In the representative 65×65 $nm^2$ field of view, the model identified 115 particles. Visually, the segmentation masks follow particle contours with good fidelity across a wide range of shapes, including rods, platelets, and quasi-spherical particles. This confirms that even a minimally tuned segmentation pipeline is sufficient for automated particle localization in heterogeneous nanoparticle ensembles. At the same time, a fraction of particles remains unsegmented. These particles primarily correspond to (1) regions with significant particle overlap and (2) areas affected by image blurring, likely due to local drift during acquisition.

This model behavior has a dual interpretation. On the one hand, it is a disadvantage: overlapping or blurred particles are not included in subsequent compositional analysis, and therefore some potentially useful information is lost. On the other hand, it can be considered as an advantage: regions with strong overlap are intrinsically problematic for compositional assignment by EDX, since the measured signal may represent a superposition of multiple particles. By not segmenting these ambiguous regions, the workflow implicitly filters out particles with poorly defined or mixed compositions, improving the reliability of the dataset used for compositional exploration.

From the segmentation results, we can estimate the particle density. With 115 particles identified within a 65 × 65 $nm^2$ region, the particle density is $\rho \sim 2.7 \cdot 10^4$ particles/μm. The experimentally observed particle density here is significantly higher than the conservative assumptions used in the Monte Carlo model in Section III (which assumed much larger effective particle radii). Therefore, even after accounting for practical limitations such as partial segmentation loss, overlap filtering, and finite grid area, the experimentally achievable particle counts are sufficient to support exploration of high-dimensional compositional spaces (≥6D under realistic grid-size constraints).

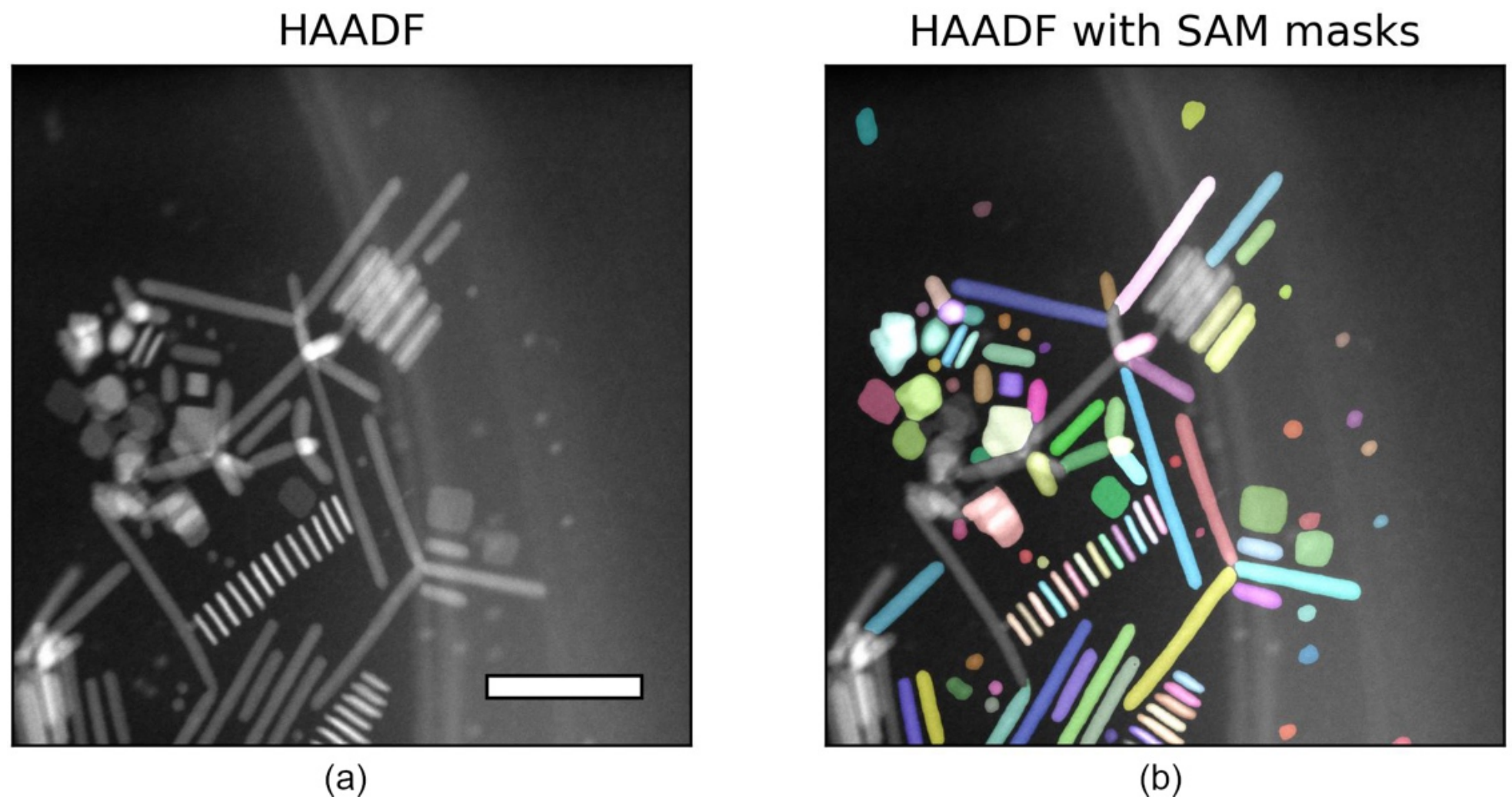

**Figure 5.** Particle identification in a heterogeneous nanoparticle library using a dummy SAM-based segmentation pipeline. (a) HAADF-STEM image, containing mixed nanostructures with varying morphologies and contrasts. (b) The same field of view overlaid with segmentation masks. The scalebar length is 7 nm.

At the next stage, EDX spectra were acquired selectively by scanning at the centers of the segmented particles (Figure 6). Such a targeted acquisition strategy significantly reduces the total measurement time, since signal does not need to be collected from empty regions of the grid. Acquisition time is a critical parameter for high-throughput characterization workflows and can easily become the primary bottleneck for large combinatorial datasets. In our experiments, spectra were recorded for each particle during 2 s at a beam current of 200 pA. Under these conditions the spectral quality is sufficient for attribution of a particle to a compositional class, but not for reliable quantitative composition estimation. Attribution based on the low-intensity spectra is inherently non-trivial due to counting noise, peak overlap, and detector/background effects. In practice, additional structural information can help resolve these ambiguities. Particle shape and local morphology provide valuable context, and establishing correlations between particle shape and composition could substantially accelerate the classification process. In the present dataset, out of 39 analyzed particles, only one particle was incorrectly attributed to the InP/ZnSe/ZnS class, while its morphology and surrounding context clearly indicate that it belongs to the CdS. Overall, even this rapid screening approach enables reliable particle discrimination, and incorporating additional structural descriptors together with slightly longer acquisition times should further improve the robustness of the classification workflow. While further improvements, including drift correction, optimized counting statistics are required, this proof-of-principle example shows that compositional assignment is technically feasible and does not represent a fundamental bottleneck for STEM-based random-library exploration.

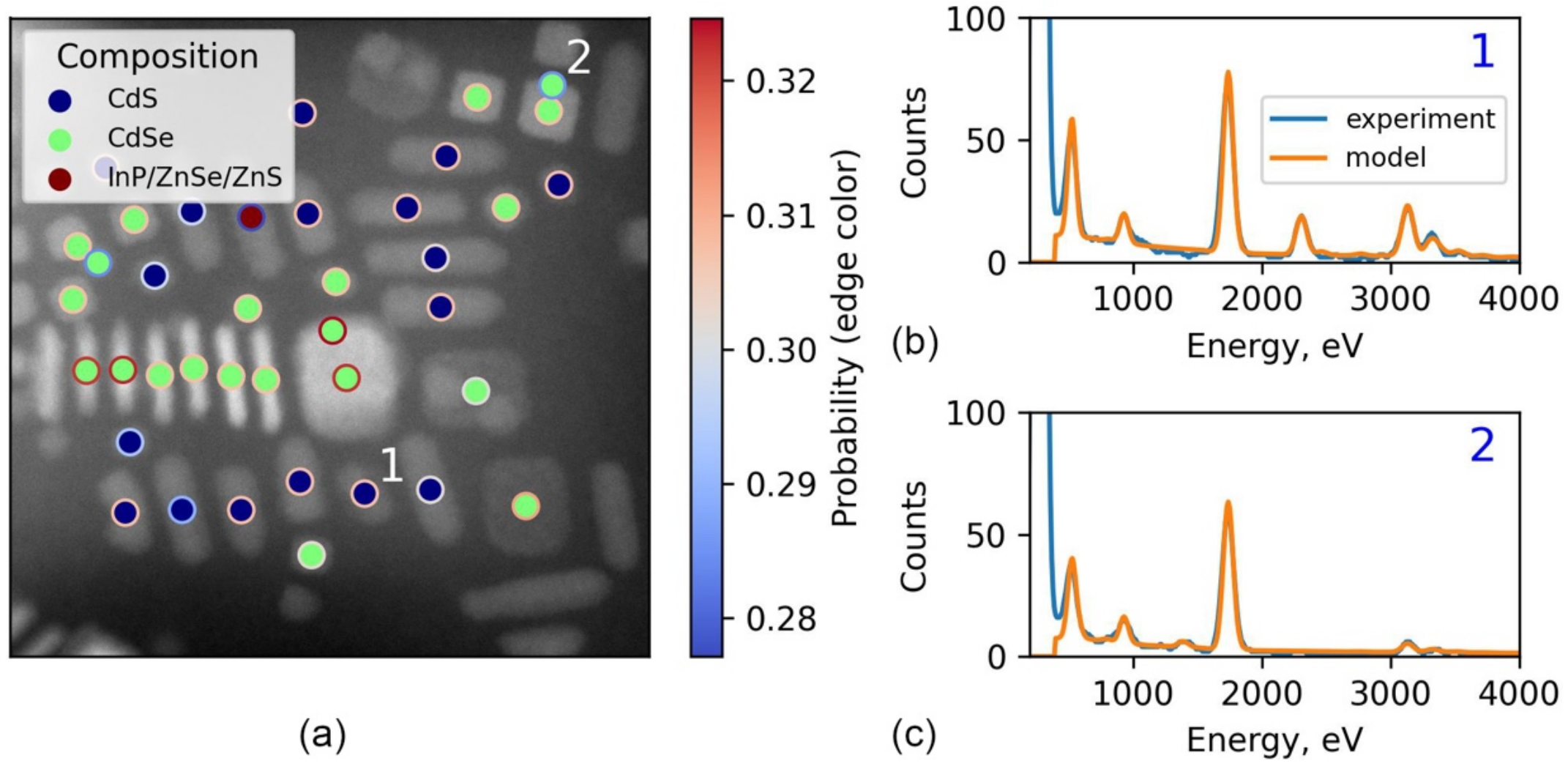

**Figure 6.** Particle classification from STEM-EDS measurements in the random nanoparticle library. (a) STEM image of the particle library with segmented particles overlaid and colored according to the attributed composition class. The colorbar scale indicates the probability assigned to the predicted class (marker edge colors). (b,c) Corresponding EDS spectra for particles **1** and **2**. The measured spectra (blue) are compared with the fitted model spectra (orange) used for composition attribution.

V.2. Computational Infrastructure for Rapid Phase Analysis

The next challenge lies in the exploration of random libraries lies in the rapid downstream interpretation of experimentally discovered phases. While modern microscopy platforms enable high-throughput acquisition of structural and compositional information, translating these observations into physically meaningful hypotheses remains a bottleneck. In practice, this stage requires access to external knowledge sources, including crystallographic databases, theoretical predictions, and prior literature.[71, 72] Efficient navigation of this information space is therefore essential for enabling open and adaptive experimental decision-making.

Among other approaches, tailored large language model (LLM) solutions provide a promising framework for integrating heterogeneous knowledge sources and generating contextual information on demand. In particular, the AtomGPTLab ecosystem[73] enables rapid access to major materials databases, including JARVIS-DFT,[74, 75] Materials Project,[76] OQMD,[77] AFLOW,[78] and Alexandria,[79] together with a suite of specialized AI models for materials analysis. These tools include DiffractGPT[80] for crystal structure identification from diffraction data, MicroscopyGPT[81] for direct analysis of microscopy images, and machine-learning models such as ALIGNN[82, 83] and SlaKoNet[84] for rapid property prediction. Through the AGAPI infrastructure,[85] these capabilities can be orchestrated within agent-based workflows that combine database retrieval, simulation tools, and machine-learning models to support accelerated materials discovery and decision-making.

As a simple proof-of-concept, we illustrate how AtomGPT can be used to obtain additional context for particles identified in the random nanoparticle library. In the first example, a query using the experimentally determined chemical formula returns candidate crystal structures together with associated information such as possible space groups and reported band gap values extracted from density functional theory databases (Figure 7a). In the second example, the model generates an approximate X-ray diffraction pattern for the candidate phase, including predicted peak positions and relative intensities for the Cu $K_\alpha$ radiation source (Figure 7b).

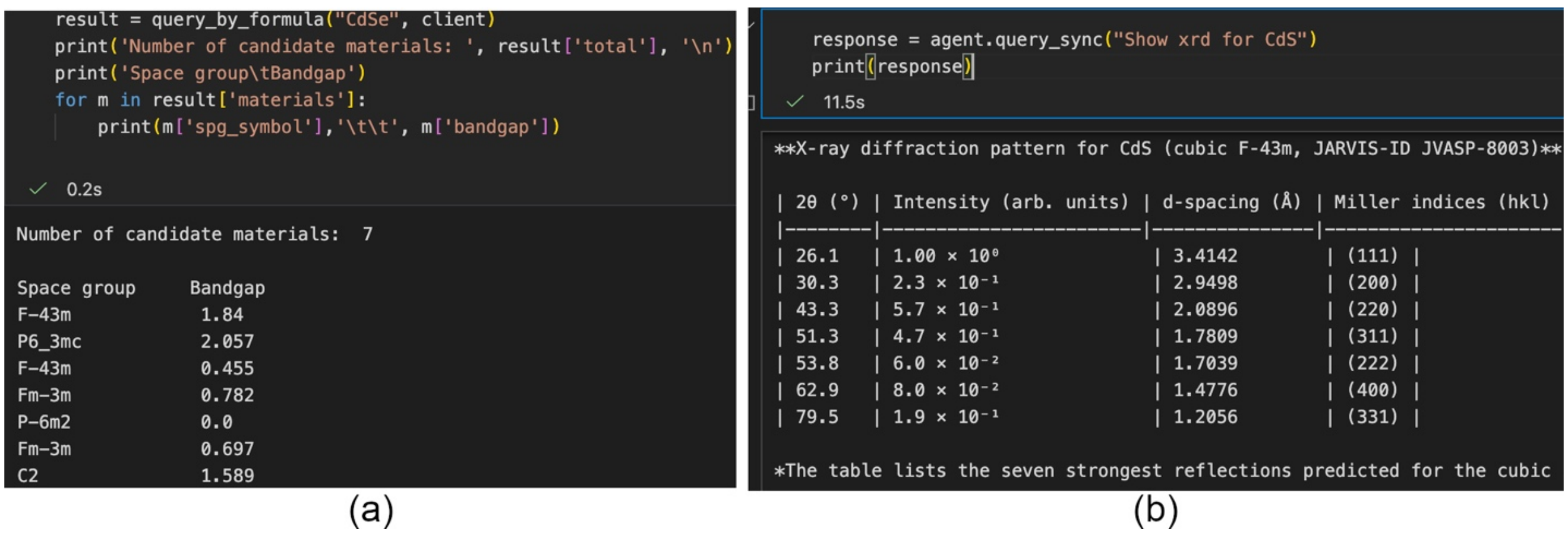

**Figure 7.** AtomGPT-assisted analysis of compositions identified in the random nanoparticle library**.** (a) Example query using the chemical formula of a particle identified by STEM-EDS. AtomGPT returns candidate crystal structures together with associated information extracted from materials databases, e.g. including space group, and reported band gap values. (b) Example generation of an approximate powder X-ray diffraction pattern for the candidate phase using Cu Kα radiation, listing predicted peak positions, relative intensities, Miller indices.

These examples represent only a minimal demonstration of the potential of the AtomGPTLab and similar ecosystems computational assistance within the automated microscopy workflow. Even such simple queries provide immediate contextual information that can guide subsequent experimental steps, including targeted structural characterization, hypothesis generation, or prioritization of promising compositions for further investigation. More broadly, integration of LLM-driven knowledge retrieval with automated experimental platforms opens a path toward more flexible and knowledge-aware optimization strategies in combinatorial materials discovery.

## VI. Summary

Machine-learning–enabled STEM is emerging as a uniquely powerful tool for materials characterization at the length scales where real functionality resides. While high-throughput synthesis from solution robotics to combinatorial thin-film deposition has, by now, become relatively mainstream, the corresponding characterization capabilities have lagged. Combinatorial spread libraries provide efficient access to quasi-ternary cross-sections of high-dimensional composition spaces, but this 2D constraint severely limits exploration strategies. Conventional photon-based characterization, dominated by laboratory X-ray diffraction and related techniques, remains intrinsically slow, and synchrotron measurements, although faster and more sensitive, carry substantial access and scheduling latencies. In this landscape, ML-enabled STEM offers a pathway to close the characterization gap by extracting maximal structural and chemical information from each electron-based measurement and by enabling intelligent, data-driven navigation of complex materials spaces.

Transitioning from photon- to electron-based characterization opens a route to much more rapid sampling of chemical space with far fewer constraints on exploration strategies. Because STEM is chemically sensitive at the nanoscale, multiple, widely separated regions of composition and phase space can coexist within a single, deliberately heterogeneous specimen, and be identified *in situ* by EDS/EELS rather than by external labels or positional encoding. The recent success of ML-driven autotuning and adaptive control suggests that instrument optimization can be carried out far more quickly and reproducibly than in fully manual workflows. Monte Carlo estimates indicate that, even with present-day instruments and conservative assumptions, effective sampling rates for structure–chemistry–vibration space can exceed those of traditional X-ray–based strategies by orders of magnitude, with substantial additional gains expected as automation and hardware improve. More broadly, this photon-to-electron shift leverages the full breadth of STEM capabilities, including atomic imaging, spectroscopy, 4D-STEM, and correlative modalities within ML-guided experimental loops, enabling flexible, high-information exploration of complex materials landscapes that would be inaccessible to conventional characterization pipelines.

Taken together, these developments suggest that rapid, ML-enabled electron microscopy can finally bring characterization onto the same footing as modern high-throughput synthesis. When solution robotics, combinatorial spread libraries, and automated reactors are coupled to STEM workflows that can interrogate many compositions and microstructures within a single specimen with minimal human tuning, the long-standing imbalance between "how fast we can make" and "how fast we can understand" begins to disappear. In this regime, synthesis and characterization become co-equal components of closed-loop discovery, with microstructure- and defect-level information feeding directly into models that propose the next experiments. We argue that this resonance between fast synthesis and fast, electron-based characterization is a necessary ingredient for a new era of materials discovery, in which exploration of complex chemical and processing spaces is limited not by bottlenecks in measurement, but by our creativity in defining objectives and rewards for autonomous experimental systems.

**Acknowledgements**

This material (BNS, SVK) is based upon work supported by the National Science Foundation under Award No. NSF 2523284. This work (AH) was supported by the U.S. Department of Energy, Office of Science, Basic Energy Sciences, Materials Sciences and Engineering Division. Scanning transmission electron microscopy characterization was performed at the Institute for Advanced Materials and Manufacturing (IAMM), University of Tennessee, Knoxville. CDL, CCB, and BMC – The experimental materials synthesis and characterization material is based upon work supported by the US Department of Energy, Office of Science, Office of Basic Energy Sciences, as part of the Energy Frontier Research Centers program: CSSAS-The Center for the Science of Synthesis Across Scales under Award Number DE-SC0019288.